\documentclass{aa}
\usepackage{graphicx}
\usepackage{amssymb}
\begin{document}
\title{On the potential of extrasolar planet transit surveys}
\author{M.  Gillon  \inst{1}  \and  F.  Courbin \inst{2} \and P.  Magain
 \inst{1} \and B.  Borguet  \inst{1}}
\offprints{M.  Gillon\\email : gillon@astro.ulg.ac.be}
\institute{Institut d'Astrophysique et de G\'eophysique,  Universit\'e
  de Li\`ege,  All\'ee du 6 Ao\^ut, 17,  Bat.  B5C, Li\`ege 1, Belgium
  \and  Laboratoire     d'Astrophysique, Ecole  Polytechnique     
  F\'ed\'erale  de  Lausanne (EPFL),    Observatoire,         CH--1290
  Sauverny, Switzerland}
\date{}
\abstract{We  analyse  the   respective   benefits and drawbacks    of
  ground--based   and    space--based     transit      surveys     for
  extrasolar planets.  Based on simple but realistic assumptions about
  the fraction of  lower main  sequence  stars harboring telluric  and
  giant planets  within  the outer limit   of  the habitable zone,  we
  predict the harvests of fictitious surveys  with three existing wide
  field    optical    and   near--IR    cameras:  the    CFHT--Megacam,
  SUBARU--Suprime and VISTA--IR.   An additional  promising instrument
  is considered, VISTA--Vis, currently under development.  The results
  are  compared  with the harvests  predicted  under  exactly the same
  assumptions, for the space missions  COROT and KEPLER.  We show that
  ground--based wide field surveys may discover more giant planets than
  space missions. However, space surveys seem to  constitute  the best 
  strategy to search for telluric planets. In this respect, the KEPLER
  mission appears 50 times more efficient than any of the ground--based
  surveys  considered  here.   KEPLER  might  even  discover  telluric
  planets in the habitable zone of their host star.  
  }
\titlerunning{On the potential of extrasolar planet transit surveys}
\maketitle
%
%%%%%%%%%%%%%%%%%%%%%%%%%%%%%%%%%%%
%
\section{Introduction}

More than  160 extrasolar   planets have been  discovered\footnote{see
  Extrasolar                      Planets                     Catalog,
  \\http://www.obspm.fr/encycl/catalog.html} since the first detection
of  a   ``Hot Jupiter''    around  the   main--sequence  star  51 Peg
(\cite{Mayor01}).  Most of these discoveries are the results of radial
velocity (RV) searches. However,  while RV searches are very efficient
at detecting massive  planets,  they do  not allow, on  their  own, to
characterize completely the orbit of the planet. Due to the degeneracy
between the mass of  the planet and the  inclination of its orbit with
respect to the line of sight (see,  \cite{Perryman}), $M$ sin$i$ is the
only measurable quantity using the RV technique alone,  where $M$ is the
mass of the planet and $i$, the inclination of  its orbit.  At the end
of 1999, a significant advance was  accomplished by the observation of
the first transit  of an extrasolar planet  discovered  through the RV
method     (\cite{Henri01};  \cite{Charbonneau01}).  With simultaneous
spectroscopic and  photometric  observations of  the star  during  the
transit, both the mass and the radius of the companion object could be
determined,     confirming  beautifully   its   planetary  nature, and
demonstrating   the great   interest of    combining  RV and   transit
observations.

Transit photometry is  not only  of interest   as a  follow up of   RV
searches, but also  as a way to find  new planets.  It may prove even
more sensitive to low  mass planets than RVs.   In fact, the method is
considered  as one of the most  promising ways of finding earth--class
planets  (\cite{Sackett01};    \cite{Schneider01}).   The OGLE  survey
(Optical  Gravitational Lensing      Experiment,     \cite{Udalski01};
\cite{Udalski02}),  a  ground--based  experiment  initially devoted  to
stellar microlensing  in the galactic  bulge, has been extended to the
search for  planetary      transits.   In  2003, Konacki   et      al. 
(\cite{Konacki01}; \cite{Konacki02}) announced the characterization of
the  first exoplanet discovered  by  transit photometry, orbiting  the
star OGLE--TR--56.   The surprisingly short period  of 1.2  day of the
companion object was confirmed by RV follow--up, a value much below the
lower end of the period distribution of planets detected by RV surveys
(\cite{Udry01}).      This  important     discovery demonstrated   the
effectiveness of  transit photometry at  detecting extrasolar planets,
and was  soon followed by  four other  discoveries  by  the same  team
(\cite{Bouchy01}; \cite{Pont01}; \cite{Konacki04}). Other surveys have
started, such as   TrES (Trans--Atlantic Exoplanet Survey)   multisite
transit  survey,  and its  first  discovery  of an exoplanet,  TrES--1
(\cite{Alonso01}). Very recently, transits of a saturnian size planet 
discovered through the RV method have been observed by the N2K Consortium 
(\cite{Sato}) around the star HD 149026, leading to a total of eight 
stars known to experience planetary transits.

With the potential of the transit method  now well established, space
missions have been proposed in the hope to discover dozens of new planets.   
Among  them are the COROT   (\cite{Rouan01}) and KEPLER (\cite{Koch1})
missions, designed to perform  high  precision photometry of  selected
fields   over  several years.   In parallel, wide    field cameras are
becoming available  on   large  ground--based  telescopes,  making  it
possible    to survey,  from  the ground,    large   numbers of  stars
simultaneously.   The aim  of   the present paper  is   to weight the
relative benefits and drawbacks of ground--based  and space surveys, by
carrying out a  {\it comparative study} of  their  efficiency based on
simple simulations.  In   particular, we investigate   whether a large
dedicated ground--based survey could be a  competitive alternative to a
high--cost space mission, or whether both approaches are complementary.

On  the basis  of realistic  assumptions,  we estimate the  harvest in
extrasolar  planets  for each type  of survey.   Our   main goal is to
compare the  merits of  each   survey  type, rather   than  predicting
absolute discovery  rates.  Such  absolute   predictions are  far  too
sensitive to  the assumptions about the  physical properties and about
the formation  process of planetary systems.   We consider only simple
distributions of planets around main sequence stars, in terms of radii
and orbital distances. These are the only relevant quantities needed to
carry out a comparative study, along with estimates of  the density of 
stars in the simulated stellar fields.

The method used to  estimate the harvests  in planets is described  in
Sect.   2.  The surveys considered and  the results of our simulations
are  presented in  Sect.   3, while  the  results and conclusions  are
summarized in Sect. 4.

\section{Description of the method}

\subsection{Strategy and field selection}
Planetary transits are  rare and the eclipse  seen in their host stars
is extremely shallow: around $10^{-2}$ mag for a Jupiter--size planet and 
$10^{-4}$ mag for an Earth-size planet in orbit around of a solar--type 
star.   The
question  of the  compromise that inevitably  has to  be  made between
field and depth is  then a critical issue  for any transit search.  In
addition, the time scales involved,  i.e.  the (expected) duration  of
the transits,  is limiting  the range of  suitable exposure  times, as
soon as  the goal is  to properly  sample the light  curves during the
transit  phase.   The maximum exposure   for   a given telescope    is
therefore mainly  dictated by the size  of  the planets one  wished to
discover   with respect to the size  of its host star.  Clearly, large
telescopes are  required in  order   to apply  as short  as   possible
exposure times and to  sample the light curves  at best. Wide field is
also of  major importance to observe many stars  simultaneously and to
compensate for the scarcity of transit events (e.g., \cite{Mallen01}).
Being aware of this general line to design surveys, we describe in the
following the  simulations used in order to   estimate the harvests of
fictitious and real planetary transit surveys.

\subsection{Suitable host stars}

\subsubsection{Density and spectral types}

For a  given  planetary    radius,  the transit  depth  is   inversely
proportional  to the square  of  the stellar  radius.  Thus, planetary
transits  will be more  easily detected in  cool dwarf stars.  In this
paper, we only consider the main--sequence stars with spectral subtypes
from $F0$ to  $M9$ as suitable candidates. We  refer to them  as Lower
Main-Sequence Stars ($LMSS$).

The galactic plane  is  the obvious  place   to look at, in  order  to
observe  simultaneously  many $LMSS$.    It is,  indeed, the  strategy
adopted  in the pioneering  work by  Mall\'en--Ornelas  et al.  (2003). 
Since the highest possible Signal--to--Noise Ratio ($SNR$) per exposure is
required, fields with  minimal reddening are  prefered.  A best choice
consists  of fields close to, but   not right in,  the galactic plane,
with galactic latitudes between  $2^{\circ}$ and $6^{\circ}$.  We have
chosen to   carry out all   our  computations for  fields  with a mean
latitude of $4^{\circ}$, where a representative extinction coefficient
is $A_V$=0.7 mag/kpc (\cite{Schlegel01}).

\begin{figure}[t!]
\centering                     
\includegraphics[width=9.0cm]{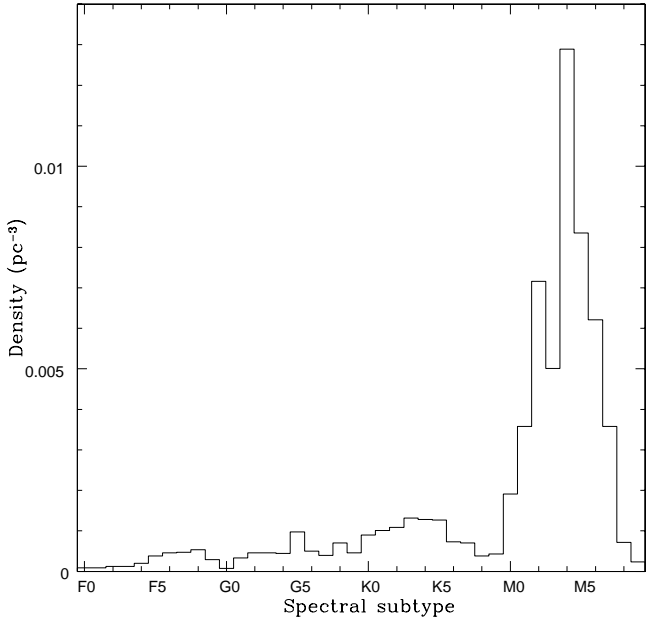}
\caption{Density of Lower Main Sequence Stars ($LMSS$)  spectral 
  subtypes in the solar neighborhood, estimated from the Gliese (1991)
  and Zakhozhaj (1979) nearby stars catalogues.}
\label{fig:sol}
\end{figure}

The projected densities of $LMSS$ at low galactic latitude are estimated
from the   Gliese  \& Jarheiss   (\cite{Gliese01})  and  the Zakhozhaj
(\cite{Zakhozhaj01}) catalogues  of  stars  in the solar  neighborhood
(see Fig.~\ref{fig:sol}).  We assume that these stellar densities also
apply to the rest   of the galactic   plane.  This is probably  a fair
approximation  as long as the galactic bulge  is avoided. The fields we are
modeling  here mainly contain stars  belonging  to the spiral arms of
our Galaxy.

\begin{figure}[t!]
\centering                     
\includegraphics[width=9.1cm]{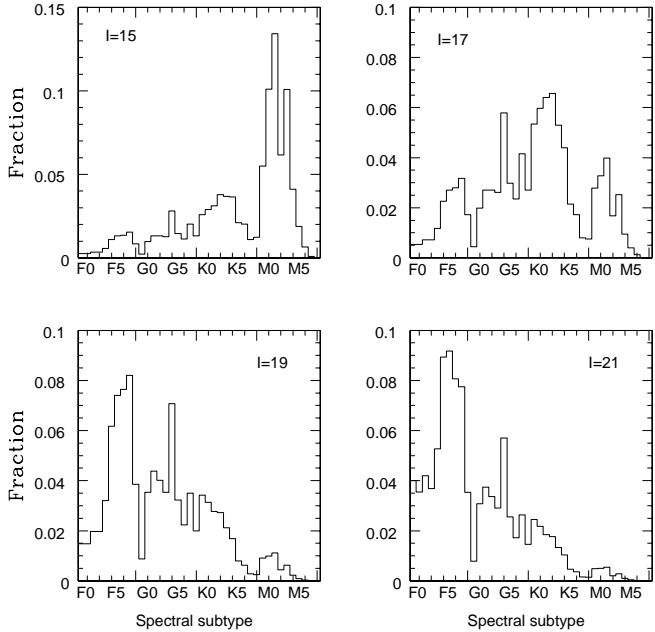}
\caption{Spectral  subtypes distribution of  $LMSS$ for 4 limiting  
  magnitudes (15, 17, 19 and 21).  An extinction coefficient
  $A_V$=0.7  mag/kpc  and a  maximum distance  $D_L$=4300 pc  are used
  (\emph{see text}). }
\label{fig:sob}
\end{figure}

\begin{figure}[t!]
\centering                     
\includegraphics[width=9.1cm]{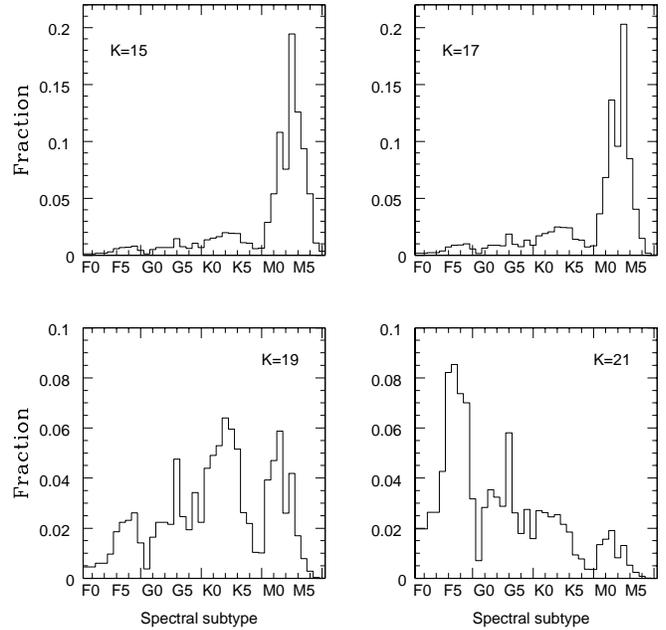}
\caption{Same Figure as Fig.~\ref{fig:sob}, but in   the
  $K$--band.   The  proportion of  very--low--mass stars,  for a given 
  limiting magnitude, is larger than in the $I$--band.}
\label{fig:sok}
\end{figure}

Under these assumptions,  the number of stars  in the field of view is
computed for every half--magnitude  bin  and for all  spectral subtypes
from $F0$ to $M9$. The volumes sampled by the  Gliese et al. (1991) and
Zakhozhaj et al.  (1979) catalogues are  scaled to the volumes of each
of the  surveys considered.  Thus, the  total number of stars observed
per half--magnitude  bin depends on the size  of the field of  view and
the filter used.  Fig.~\ref{fig:sob} shows the distribution of $LMSS$,
computed for all  subspectral types and for 4 limiting magnitudes in
the  $I$--band.  Going to  the  infrared allows   to  observe a  larger
fraction of very--low--mass stars, as is shown in Fig.~\ref{fig:sok}.

\subsubsection{Blends and binarity}

All   our calculations are  exclusively  made  for $LMSS$  and  do not
include the effect of blends, for two reasons:

\begin{enumerate}
  
\item   Introducing all  spectral types will   increase  the number of
  potentially  interesting   transits.   It   will also increase   the
  crowding of the  fields.  Since stars other  than $LMSS$ have larger
  diameters, the net result of the competition between the two effects
  is likely a reduction of the true observable number of transits. Our
  results are therefore upper limits on the number of transits seen by
  each survey.  However, since our goal is a comparative study between
  surveys, rather than an attempt  to predict absolute transit counts,
  leaving aside the spectral types other than $F0$ to $M9$ does not
  affect our conclusions.\\

\item More blends also  mean an increased number  of false detections. 
  For Jovian--mass planets, a large  part of   these false   detections 
  can  be rejected through a fine--tuned transit light--curve analysis   
  or with  follow--up  RV observations (\cite{Bouchy2005}). In some cases,  
  deciding whether an eclipse is due to a genuine planet or not can be 
  very tricky (see e.g.  \cite{Mandushev2005}). The difficulty increases
  drastically in the case of a telluric planet. 
  It does not only depend on the planet type, but also on the method
  of analysis used  to  build the  stellar light   curves and on   the
  transit detection method  and  selection criteria: methods based  on
  aperture photometry  will be heavily affected  by seeing and blends,
  while differential  imaging,  PSF  fitting, or   image deconvolution
  behave very  differently with  respect  to blends and to PSF  mismatch. 
  Finally,  prior  knowledge  on  the  exact field  geometry    can be
  introduced as well in the analysis, e.g.,  by targetting fields with
  archived  HST   images  used  as reference  maps   to carry  out the
  photometry.  The degree of additional complication due to blends is
  such   that it falls well  out  of the scope  of  the present paper. 
  Again, only relative estimates between surveys are aimed at.

\end{enumerate}

Very little is known about the formation of planets in binary systems. 
Furthermore, detecting transits in close binary systems is possible but 
can be difficult, because of the decrease of the transit depth.  We   have 
chosen to   adopt an arbitrary fraction  of 50\% of  binary $LMSS$ and to
do not allow a transit detection around them.  The exact  value  of this 
fraction   has no  influence  on the result of  our comparative study.  
Our whole simulations are designed to  predict the maximum number of
planets  that can  potentially be discovered with  each  type of  survey, 
with  a slight and  arbitrary correction for binarity.

\subsubsection{Distance}

Among our  assumptions is the maximum distance above which the density
of $LMSS$ may not be  valid.   As we consider   fields with a low  but
non--zero galactic latitude, this maximum distance can be computed from
the thickness of  the galactic disk.   Considering a thin  disk with a
thickness of 600 pc, positioning the Sun at mid  height, and using an
average galactic latitude    of   $4^{\circ}$ for  all  the    surveys
considered  in this work,  the limiting  distance $D_L$ is simply:

\begin{equation}\label{eq:a9}   
D_L=\left(\frac{600}{2}\right)\times\left(\frac{1}{\sin{4^{\circ}}}\right) \approx 4300 \textrm{ pc}
\end{equation} 

In the following we will consider that the density of $LMSS$ drops
to zero beyond the distance $D_L$.  The limiting magnitudes imposed to
us by the telescopes and cameras chosen in  this work for ground--based
surveys are  well above the  limiting  magnitude due to  pure distance
effects.  The  magnitude range is therefore  limited by distance
rather  than  by  instrumental depth.   For space--based  surveys,  the
instrumental depth is the limiting factor.

\subsection{Estimated number of planets in the field}

Once the number of target stars has  been estimated, one also needs to
    make  assumptions about  the  average number  of planets hosted by
$LMSS$, and on the distribution of their radii and periods. We consider
six types of planets:

\begin{itemize}
 
\item{{\bf Very Hot Jupiters ($VHJ$)}: we define these objects as giant
    planets with a maximum  period of 3 days  and an arbitrary minimum
    period of 1 day. On the basis of  the results of \cite{Gaudi5}, we
    assume that 0.2\% of $LMSS$ may host a $VHJ$. We use a rough radii
    distribution represented by  a sigmoid function, interpolated from
    the   radii distribution observed in   our  own solar system  (see
    Fig.~\ref{fig:ges}).  We  define  the $VHJ$--Zone as  the range of
    distances to   the  host star   corresponding to  orbital  periods
    between 1 and 3 days. We assume that the periods of the $VHJ$ are
    uniformly distributed in the $VHJ$--Zone.}\\
 
\item{{\bf Hot Jupiters  ($HJ$)}: are defined as  giant planets  with a
    period in between 3 and 9 days.  Following Gaudi et al. (2005), we
    adopt a value  of 1\% for the  fraction of $LMSS$ orbited  by such
    objects.  We  use the same radii  distribution as   for the $VHJ$. 
    We define the  $HJ$--Zone as  the region around  a  star where the
    period of a planet is between 3 and 9 days. As noticed by Gaudi et
    al. (2005), about  half of the  $HJ$ have a  period lower than 3.5
    days.   We thus  consider   that half  of  the $HJ$   have periods
    uniformly distributed in the range 3--3.5 days,  and that the other
    half have periods uniformly distributed in  the range 3.5--9 days. 
    $HJ$ and $VHJ$ forms the  group of the Close--in Extrasolar  Giant
    Planets ($CEGP$).  We shall denote them $(V)HJ$ when
    considered as a single group.}\\
 
\item{{\bf Giant planets in the Habitable Zone ($GHZ$)}: giant planets
    in the  zone identified by Kasting et  al.  (1993) as suitable for
    life.  We use the  limits established by Kasting  et al.   (1993)
    for the   Zero Age  Main  Sequence $HZ$  ($ZAMS$  $HZ$)  and their
    "intermediate"  habitability  criteria.   Based   on the  measured
    fraction  of extrasolar planets found  in the $HZ$, we assume that
    3\%  of the $LMSS$  harbor such a  planet.   We use the same radii
    distribution  for these  objects as for  the $HJ$  and  $VHJ$.  We
    consider that the  periods of the  $GHZ$ are uniformly distributed
    across the $HZ$.}\\

\begin{figure}[t!]
\centering                     
\includegraphics[width=9.1cm]{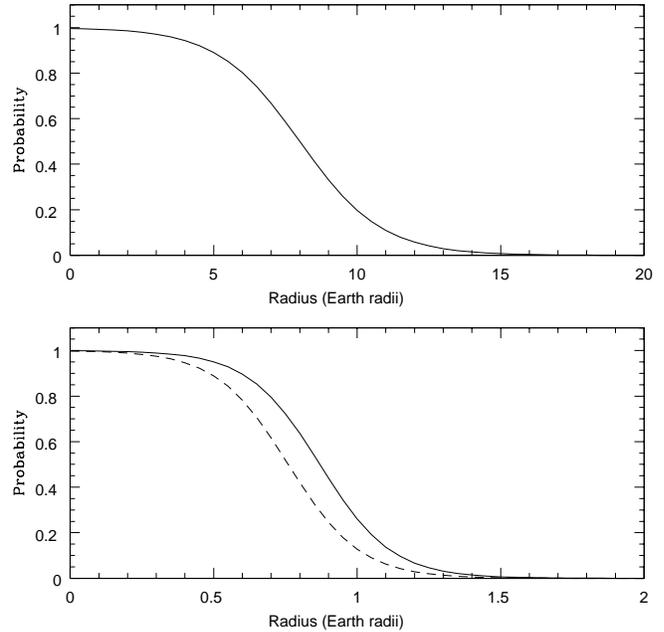}
\caption{\emph{Top:} Radii    distribution     of    giant      planets 
  interpolated from the  four giant planets  in our Solar  System. The
  y--axis gives the probability that a  planet has a radius larger than
  the value on the  x--axis,   in earth radii.   \emph{Bottom:}  Radii
  distribution  of $THZ$    (solid  line) and   $TIZ$ (dotted   line),
  interpolated from the   simulations of  the formation of   telluric
  planets by Chambers (2001). }
\label{fig:ges}
\end{figure}

\item{{\bf Giant  planets  in the   Intermediate Zone  ($GIZ$)}:   the
    Intermediate Zone is located between  the $HZ$ and the $HJ$--Zone. 
    We assume that the fraction of $LMSS$ with a  $GIZ$ is the same as
    in the $HZ$,  i.e., 3\%.  The same radii  distribution is used for
    $GIZ$  as   for the other  types of giants  planets considered. We
    consider that the periods of the $GIZ$ are
    distributed uniformly in the Intermediate Zone.}\\
  
\item{{\bf Telluric planets in the Habitable Zone ($THZ$)}: nothing is
    known about   telluric planets outside our   own Solar System.  We
    base  our   estimates   on  the   terrestrial   planets  formation
    simulations of Chambers  (\cite{Chambers01}).  Using these results
    and    the  assumption  that  all   $LMSS$   are  surrounded  by a
    protoplanetary disk  at the  beginning  of their life  (which will
    lead to  the formation of telluric planets)  we obtain 1.78 as the
    average  number  of terrestrial  planets in  the $HZ$ per  $LMSS$. 
    This number does not take into account the fact that, for a poorly
    constrained fraction  of the stars,  one or  more giant planet may
    migrate inwards over the outer limit of  the $HZ$, hence affecting
    the orbits  of the  telluric planets  and maybe accreting  them or
    expulsing them on a hyperbolic orbit. We take this fraction as the
    sum of the fractions  of $VHJ$, $HJ$, $GHZ$  and $GIZ$, leading to
    7\% of the $LMSS$ hosting a giant planet inside the outer limit of
    the $HZ$.  Assuming that  no terrestrial planet can  survive under
    these conditions, we obtain $1.78 \times (1-0.07) \approx 1.66$ as
    the average number of terrestrial planets  in the $HZ$ per $LMSS$. 
    We also    use the    results  of  Chambers   (\cite{Chambers01}),
    interpolated with  a sigmoid  function (see Fig.~\ref{fig:ges}) to
    estimate the radii distribution  of telluric planets  in the $HZ$. 
    We consider  that    the periods of   the   $THZ$ are  distributed
    uniformly
    in the $HZ$.}\\
     
\item{{\bf Telluric  planets in the  Intermediate Zone ($TIZ$)}: using
    the results of Chambers (\cite{Chambers01}) and correcting for the
    fraction of $LMSS$ having a giant planet inside the outer limit of
    the $IZ$  (4\%),  we estimate the  average number  of terrestrial
    planets in the $IZ$ to be 1.51 per $LMSS$. The results of Chambers
    are  again used  to estimate the  radii   distribution of telluric
    planets in the $IZ$ (Fig.~\ref{fig:ges}).  We consider that the
    periods of the $TIZ$ are distributed uniformly in the $IZ$.}\\

\end{itemize}

The  planets outside the $HZ$ are  not considered  here, as their long
period and  their low geometric probability  of producing a detectable
transit make  them  very poor  candidates.  We   also do  not consider
telluric planets  with a period shorter than  9 days, because they are
not predicted by the simulations of  Chambers and because there is  no
such planet in our own solar system.

Our  assumptions about  giant  planets  are empirical,  and  based  on
previous surveys that  are certainly  biased  towards a given type  of
planet and  orbit  and   still  suffer  from  low number  statistics.  
Furthermore, we assume that the average expected number of planets for
the  four zones we defined earlier are the same  for all spectral subtypes
from  $F0$ to $M9$. This assumption does not have a strong influence on 
the results given our already rough knowledge of planetary formation, 
although  a metallicity
dependence on exoplanets population has been emphasized (see e.g.
\cite{Santos}) and low--mass stars tend to be older on average  and have 
a lower metallicity.
The  predicted harvests should nevertheless be of
the right order of magnitude and are certainly adequate for comparison
purposes.

The assumptions  made on  telluric    planets are  based on a   purely
theoretical work.  We do not  take into account the possible existence
of giant planet cores formed in the outer  region of the disk and that
later  migrated  inwards.   These   objects, without  a massive  primary
atmosphere, could consist of  large planets of  pure rock and  ice. We
prefer to  remain   conservative and to  consider  exclusively planets
formed in the more standard scenario thought to be responsible for the
formation of planetary systems such our own.

The known weaknesses of the  assumptions used  in our simulations  are
not a critical issue for our purpose, which is to compare the relative
strengths and  weaknesses of different transit  surveys rather than to
predict actual discovery rates for a given survey.  Our results may be
scaled up and down, but we believe that they remain correct as long as
the aim is to carry out comparative studies.

\subsection{Parameters of the instruments chosen for the fictitious 
ground-based surveys}

Transit   searches must combine large field    of view, depth and good
temporal sampling.   Although the  range of possible  telescope/camera
combinations  is broad, we have  selected four instruments that can be
considered  as  representative  of  the   present  or soon   available
astronomical facilities.  Three   of these instruments  are already in
use: CFHT--Megacam, Subaru--Suprime, and Vista--IR.  The latter one is
only       proposed:        Vista--Vis\footnote{see     VISTA      web
  site,\\http://www.vista.ac.uk} (see Table~\ref{tab:vur}).

\begin{table}
\centering
\caption{Instruments considered in our fictitious ground--based surveys.}
\begin{tabular}{lccr}
\hline\hline Instr. & Telescope      & Field         & Location\\
                    &  diameter  & (sq. degrees) &   \\
\hline
Megacam          & 3.6 m  & 0.90 &   Hawaii (CFHT)   \\
Suprime          & 8.3 m  & 0.26 &   Hawaii (Subaru) \\
Vista--IR        & 4.0 m  & 1.00 &   Chile (VLT)     \\
Vista--Vis       & 4.0 m  & 2.25 &   Chile (VLT)     \\
\hline
\end{tabular}
\label{tab:vur}
\end{table}

For each  instrument, we have   estimated the $SNR$ and the  saturation
magnitude for a range of exposure times, for a fixed airmass (1.6) and
typical seeing (1 arcsec). These parameters are computed for different
filters, using the   Exposure Time  Calculators  (ETC)  available  for
Subaru   and the   CFHT.  The  bright   cut applied  to the  magnitude
distribution of  host stars corresponds to  saturation time.  The faint
cut corresponds to the  magnitude of stars that  have a $SNR<10$, where
a shallow transit would not be detected.

For Vista--Vis, we  used the information  given in  the online  ETC of
the EMMI camera.  This ESO  instrument is mounted on  the 3.5 m NTT at
La  Silla (Chile) and  has a throughput  similar to  Vista, which will
also be a 4m--class telescope.  For Vista--IR,  we use the  information
given on the online ETC of the SOFI IR camera of the ESO 3.5m NTT.

All our estimated  $SNR$ are  corrected for the  extra source of noise
introduced by stellar variability, following the formula:

\begin{equation}\label{1} 
SNR_2 = \frac{SNR_1}{\sqrt{1 +(\sigma_{\ast} SNR_1)^2}},
\end{equation} 

where $SNR_1$   and  $SNR_2$ are   the $SNR$  before and  after taking
$\sigma_{\ast}$ into account,  the  standard deviation of the  stellar
variability. We adopt $\sigma_{\ast} =  100$ ppm as estimated from the
solar variability within a frequency interval that matches our adopted
range of transit durations (\cite{Borde2}). We consider this same value 
for every $LMSS$ spectral subtypes.

\subsection{Computation of the harvests and detection criteria}

Under  all  the above   assumptions, the expected  numbers of  transit
detections are estimated as follows.

For every spectral subtype  F0 to M9,  we compute 100  circular orbits
for each  of the four  zones considered, distributed as  mentioned in
Section 2.3.  The geometric transit probability for  an orbit is given
by

\begin{equation}\label{2}
P_{tr}= \frac{R_{\ast}}{a},
\end{equation} 

where $R_{\ast}$ is the radius of the star and $a$ the semi--major axis
of the planet orbit.  For every orbit  and for a range of inclinations
and planetary radii, we compute the total crossing--time of the transit
and the  duration of the flat part  of the light  curve, following the
calculation by Mall\'en--Ornelas et  al.  (2001).  The results are then
averaged on the inclinations and radii.

A window function is   computed for each  survey,  based on the  total
number of nights in the campaign and the visibility  of the field each
night, in the case of  ground--based surveys.  To take weather  effects
into account,  we assume that  for 1 night  out of 10, no observations
are  taken  at all.   In  addition, 10 chunks of   1 hour are randomly
removed from the 9 remaining  nights to account for technical problem,
clouds, or unexpected overheads.

We then compute a probability ${P}_{visN}$ that  a transit is observed
$N$ times for a specific orbit and a specific $LMSS$.  We compute this
probability for $N$  varying from 1 to $X$,  where $X$ is  the maximum
number of transits we could observe for the same star during the whole
observing season.  As the shortest period  considered is 1 day, $X$ is
simply the duration of the survey in days.

For a specific spectral type and  a specific distance, the probability
$P_{obsN}$ that a  transit occurs $N$  times during the survey  and is
observable is $P_{obsN}$ = $P_{tr} \times {P}_{visN}$. 

We then consider  that the observed dimming  of  a light curve  can be
attributed to a  genuine planet  only if at  least three  eclipses are
detected.  We define $k$ as the number of transits observed during the
survey and impose that the $SNR$ of the light curve, integrated on the
duration of the flat part of the $k$ transits is at least $\beta$ time
greater than the inverse of transit depth, i.e.

\begin{equation} \label{4}
SNR  \geq \frac{\beta}{\sqrt{k}  }  \left( \frac{R_\ast}{R_p}  \right)
^{2},
\end{equation} 

where $R_p$ is the planetary radius and $R_{\ast}$ the radius of the
host star.

In other words, the significance of  planetary  transits in a stellar
light curve is $\beta \times \sigma$.

We have chosen to adopt $\beta=9$. This value is high enough to reject
the majority of the statistical artifacts.  Is is also the one adopted
by  the OGLE--III team  (Udalski,   private communication).  The  value
$\beta=7$ is also tested, as it is  used in other transit studies such
as in Bord\'e et al. (2003) for the COROT mission.

For a specific planet transiting $k$ times during  the survey, and for
a star of a given magnitude and a  given radius $R_\ast$, equation (4)
allows a computation of the minimum planetary radius needed  to get a $SNR$
high enough  to allow a    statistically significant detection.    The
relevant radii distribution then leads to a determination of the probability
that the planet has a radius  at least equal  to the minimum radius of
detection, i.e. the fraction of planets which  would produce a dimming
strong enough to be detected.

For every half   magnitude bin  and  for  every  zone considered,  the
results are averaged over 100  orbits and multiplied  by the number of
stars of each  spectral type present in each   half--magnitude bin.  We
then multiply  by  the fraction  of stars  expected  to host  a planet
(giant  or telluric) in the  zone considered, leading  to the expected
number of planet detections.  As  a final step,  we sum the numbers of
planets for each half--magnitude  bin and spectral subtype, and  obtain
the total number of detections for each planet type.

\section{Results}

\subsection{Analysis of existing surveys : OGLE--III and EXPLORE--I}
We  have tested our simulations  on two existing ground--based surveys,
EXPLORE--I  (\cite{Mallen01})    and     OGLE--III  (\cite{Udalski01};
\cite{Udalski02}; \cite{Udalski03}).   The   goal was to  compare  our
predictions to  the actual results of  these surveys and  to check the
validity of our assumptions. Our  main goal remains the comparison of
existing ground--based    surveys  to future   fictitious  ground--based
surveys and to space missions.

\begin{table}[t!]
\centering
\caption{Results predicted for  EXPLORE--I. In the first  line, we adopt
${P}_{vis2}=0.1$  for $(V)HJ$ and  ${P}_{vis2}=0$ for  the other
types of planets. The 
second  and  third   lines involve a ${P}_{visN}$  computed  assuming  good 
weather conditions   (see   text). The values in  brackets   are   for 
$\beta = 7$, and the others correspond to $\beta=9$.}
\begin{tabular}{ccc}
\hline\hline 
Weather conditions & $VHJ$ & $HJ$\\
\hline  
Real conditions ($k = 2$) & 0.4 (0.5) & 1.2 (1.5)\\
Good weather ($k \ge 2$) & 3.4 (4.2) & 2.8 (3.4)\\
Good weather ($k \ge 3$) & 2.2 (2.7) & 0.7 (0.9) \\
\hline
\end{tabular}
\label{tab:exp}
\end{table}

\subsection{EXPLORE--I.}  The  Extrasolar  Planet Occultation Research
  search lasted 11  nights, at the CTIO 4m  telescope, in the $I$--band
  with the  mosaic II  camera.  It   concentrated  on one  single 0.36
  deg$^{2}$ field near   the galactic plane  ($l$ = $-27.8\,^{\circ}$,
  $b$=$-2.7\,^{\circ}$)   containing   $\sim$  100,000  stars  down to
  $I$=18.2 and $\sim$ 350,000 stars    down to $I$=21.0.  Poor weather
  conditions  led to  a    degraded  window function, affecting    the
  probability ${P}_{vis2}$ to detect two  transits of the same planet. 
  ${P}_{vis2}$  was estimated as 0.1 for  $(V)HJ$ by  the EXPLORE team
  (Mall\'en--Ornelas, private communication).
  
  The typical exposure time was 60 seconds, and the detector read time
  plus overhead amounted to 101 seconds.   The analysis of the data is
  not completed, and no final detection  criterion has been decided up
  to  now (Mall\'en--Ornelas,   private  communication).  No  exoplanet
  transit has been discovered so far.
  
  Following   the  simulations   presented    in  this  paper,    with
  ${P}_{vis2}=0.1$ for  $(V)HJ$  (and ${P}_{visN>2}$ set   to  0, i.e. 
  assuming that the probability  to detect three  transits of the same
  planet is negligible),      we compute the   expected  harvest   for
  EXPLORE--I, given its true weather conditions.   We also compute the
  harvest  for    good weather    conditions,  i.e.    without  fixing
  ${P}_{visN}$ but estimating it  according to the method described in
  Section  2.5.  Our results for the  two detection criteria ($\beta =
  9$ and $\beta = 7$) are  presented in Table~\ref{tab:exp}.
  
\begin{figure}
\centering                     
\includegraphics[width=9.1cm]{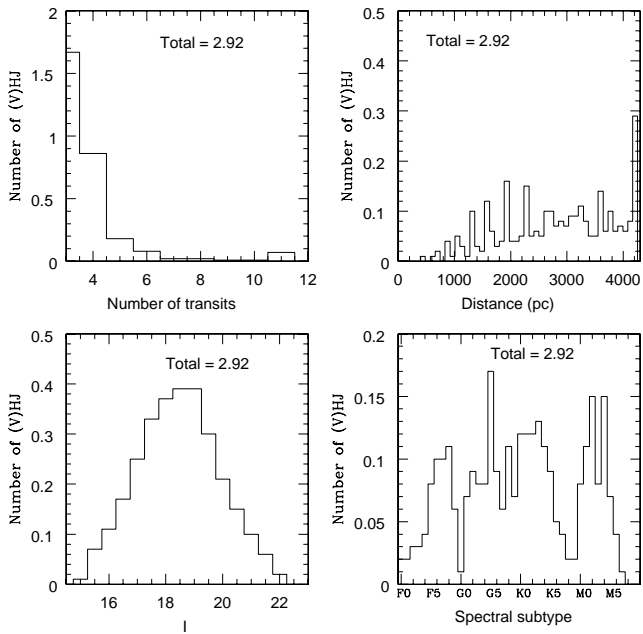}
\caption{Distribution of the expected number of $(V)HJ$ for EXPLORE--I   
(with   good   weather conditions) as a  function  of  the  number  of 
transits observed (\emph{upper left}), of  the  distance  (\emph{upper 
right}), the  magnitude (\emph{bottom left}) and  the spectral subtype 
(\emph{bottom right}) of the host  star.  The  irregular  shape of the 
histogram as a function of distance is an artificial effect caused  by 
the rounding off of the stellar magnitudes to the nearest integer   or
half--integer.}
\label{fig:exp}
\end{figure}

We  present in  Fig.~\ref{fig:exp} the  magnitude  distribution of the
host  stars $(V)HJ$ expected according   to our simulation under  good
weather conditions and $\beta = 9$.  Fig.~\ref{fig:exp} also shows the
distribution of the   distances, of the spectral  subtypes  and of the
number of transits observed for the expected discoveries.
  
Mall\'en--Ornelas et al.  (2001) expected to find 1 transiting $(V)HJ$.
Although this is very  low number statistics,  this  value is in  good
agreement  with our result  for ${P}_{vis2}=0.1$.   The absence of any
detection   is also compatible with  our  estimates. We point out that
good weather conditions  could  have led to the   discovery of a   few
$(V)HJ$, showing   that   searches using 4m--class  telescopes    and a
wide-field camera are promising strategies.  EXPLORE--I lasted only 11
nights.  A simple scaling of the results for a 3--months survey shows
that a harvest of several  tens  of $(V)HJ$ is  not unrealistic.\\

\subsection{OGLE--III}

The Optical Gravitational Lensing Experiment  entered its third phase,
OGLE--III,  in   June 2001.    It   took place   at  the Las  Campanas
Observatory, Chile, using the 1.3m Warsaw telescope  and the 8k MOSAIC
camera,  with a  total   field   of  view  of  0.34  deg$^{2}$.    All
observations were made through the $I$ filter.  Four surveys have been
carried out so far :

\begin{itemize}
\item OGLE--III--1 (June 12 to  July 28, 2001).   More than 800 images
  of  three   fields  in  the direction  of   the  galactic bulge were
  collected within 32  nights. The exposure time  was 120 seconds, and
  each field was observed every 12 minutes.
\item OGLE--III--2  (February  17 to  May 22,  2002).  More than  1100
  images of three fields located in the Carina  region of the galactic
  disk were collected in 76 nights. The exposure time was 180 seconds,
  and the temporal resolution was about 15 minutes.
\item OGLE--III--3 (February  12 to March  26, 2003).  The photometric
  data were collected  during 39 nights  spanning  the 43 days  of the
  survey. Three fields of the galactic disk  were observed with a time
  resolution of about 15 minutes. The exposure time was 180 seconds.
\item OGLE--III--4. Starting on March 25,  2003, this survey collected
  its   main  photometric   material   until  middle  May   2003,  but
  observations  were also  collected  until July 25,  2003 with sparse
  temporal sampling.  Three fields  of the galactic disk were observed
  with the same  exposure  time and  resolution  as the two   previous
  surveys.
\end{itemize} 

Using the  observation windows and  instrumental parameters of these 4
surveys (A.  Udalski,  private  communication), we have   computed the
expected   harvests    for the     two   detection     criteria   (see
Table~\ref{tab:ogl}).  OGLE--III--1 and   OGLE--III--2  have  yield  5
genuine extrasolar planets so far.  OGLE--III--3 and OGLE--III--4 have
discovered 40 transiting  companions,  but follow-up spectroscopy  has
not given the final harvest  yet.   We note that, although  predicting
absolute planet  counts is not our main  goal,  our OGLE--III expected
harvests are in very  good agreement with the  actual ones.  Our  work
hypotheses thus allow to compute realistic harvests in the case of $(V)HJ$.

\begin{figure}[t!]
\centering                     
\includegraphics[width=9.1cm]{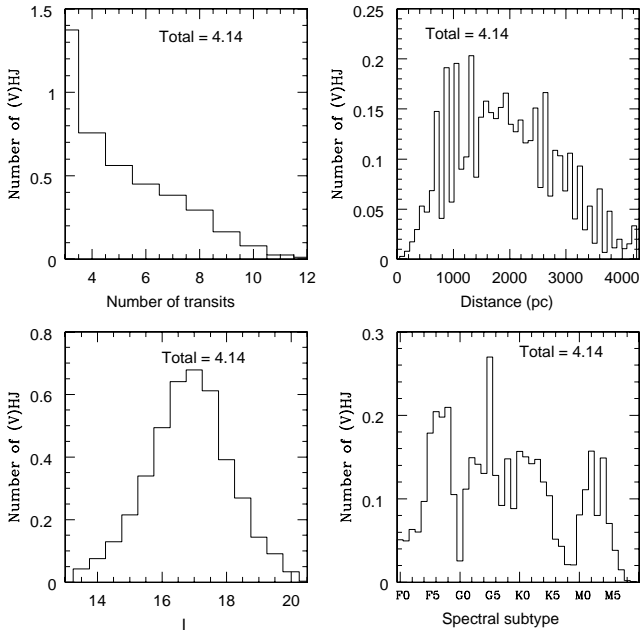}
\caption{Distribution of the 
  $(V)HJ$ discoveries for OGLE--III--2 as  a function of the number of
  observed  transits  (\emph{upper  left}),     distance  (\emph{upper
    right}),  magnitude  (\emph{bottom left})   and of  the  host star
  spectral subtype (\emph{bottom right}).}
\label{fig:ogb}
\end{figure}

Fig.~\ref{fig:ogb} shows  the distribution of  the $(V)HJ$ discoveries
predicted by  our simulation for   OGLE--III--2 as a  function of  the
characteristics   of  the host star  and    of the number  of observed
transits.   Most detections occur for  stars  located between 1500 and
2500 pc from the Sun, well matching the range of distances for the real
planets  OGLE-TR-111,  OGLE--TR--113   and OGLE--TR--132, i.e.,  about
1500 pc (\cite{Schneider02}). We also note that our predicted ratio of
$VHJ$  to  $HJ$ is  in  good  agreement  with  the  observed  one (see
Table~\ref{tab:ogl}).   Future  planet harvests  with OGLE--III--3 and
OGLE--III--4 shall allow to better  constraint the absolute counts  of
$VHJ$ around $LMSS$.

\begin{table}
\centering
\caption{Results  predicted  by our  simulations for OGLE-III surveys,
compared  to   the   actual    harvests. The values are given 
 for $\beta=9$ and the values in parenthesis
correspond to $\beta = 7$. No actual harvest is available yetfor
 OGLE--III--3 and --4 yet.} 
\begin{tabular}{ccccc}
\hline\hline 
Survey & Planet & Predicted &  Actual harvest \\
\hline  
OGLE--III--1& $VHJ$ & 1.6 (2.2) & 1\\
        & $HJ$ & 0.7 (0.9) & 1\\
OGLE--III--2 & $VHJ$ & 2.7 (3.6) & 2 \\
        & $HJ$ & 1.5 (2.0)& 1\\
OGLE--III--3 & $VHJ$ & 2.1 (2.8) & $-$ \\
        & $HJ$ & 0.8 (1.1)& $-$ \\
OGLE--III--4 & $VHJ$ & 2.6 (3.6) & $-$ \\
        & $HJ$ & 1.8 (2.5) & $-$ \\
\hline
\end{tabular}
\label{tab:ogl}
\end{table}

\subsection{Analysis of fictitious surveys}
\subsubsection{Setting the parameters of the surveys}

All the  surveys we consider in  the following  use red filters, i.e.,
redder than the $V$-band.   This choice is  motivated by the fact that
limb darkening and atmospheric and  galactic absorptions are minimized
in the  red. It also  maximizes the number  of available $LMSS$, hence
increasing the probability of detecting extrasolar planets.  Extending
the above argument from   the visible to  the near-infrared,   we have
included surveys using  the $J$,  $H$ and  $K$ filters,  on VISTA--IR. 
However, we have  not included the  longer wavelength filters  such as
$L$ or $M$,  because the quantum efficiency  of the detectors  is much
lower with present-day instrumentation.

The choice of   the exposure time  is  a  critical  issue  for surveys
carried out in the  visible.  Long exposure  times lead to high $SNR$,
but also to a  far  too large  number of  saturated stars in  galactic
fields.  Increasing the   exposure times also increases crowding,  and
consequently decreases    the $SNR$ for   any contaminated   star.  As
mentioned in  Sec.  2, effects of  crowding are not taken into account
here, and will be the subject of a forthcoming paper.   If the goal of
the survey is to discover  planets as small  as possible, a compromise
has to be found between the duration of the exposure time plus readout
time of  the CCD, the desired  $SNR$ in each  individual exposure and
the number  of sampling points  during an eclipse.   We have estimated
the exposure time leading to the largest  number of unsaturated $LMSS$
with sufficient $SNR$   and chosen to  test three  different  exposure
times close  to this "optimal" value  for each optical instrument. The
actual readout time of  each instrument is also  taken into account in
our calculations.

The situation  is simpler for  near--IR instruments, since the readout
time is  negligible.  The harvest is  thus largely  insensitive to the
adopted exposure time and we have adopted the exposure time leading to
the largest number of unsaturated $LMSS$.

Finally, we have tested surveys of four different durations: 30, 60,  
120, and 180 nights. 

We have checked that there exist fields in the  galactic disk that can
be observed continuously during  six  months, either from Paranal   or
Mauna  Kea.   However, their visibility  is reduced   to about 4 hours
during the first  and the last month of  a six months period.  We have
used the seasonal observability of typical fields of the galactic disk
to  build our   window  function and  we have    also investigated the
interest of extended surveys, carried  out over several years (between
1 and 4 years), similarly to space missions such  as COROT and KEPLER,
but with non-continuous visibility.

The  predicted  harvests for the different   surveys are summarized in
Table~\ref{tab:ret}. The columns in all Tables  are labeled as exposed
in Section~2.

\begin{table*}
\centering
\caption{Results obtained for fictitious surveys. ET  =  exposure  + 
readout time. The values are for $\beta = 9$. The values given between
parenthesis correspond to $\beta=7$. }
\begin{tabular}{ccccccccccc}
\hline\hline Instr.& Band &Years/Nights& ET
($s$)&$VHJ$&$HJ$&$GIZ$&$GHZ$&$TIZ$&$THZ$\\
\hline
CFHT& $i'$ & 1/30 & 72 & 13.0(15.2)& 14.6(17.6) & 0.1(0.1) & $-$ & $-$ & $-$\\
& & & 102 & 15.3(17.8) & 17.5(20.7) & 0.1(0.2) & $-$ & $-$ & $-$\\
& & & 162 & 16.9(19.6) & 19.4(23.0) & 0.2(0.2) & $-$ & $-$ & $-$\\
& $r'$ & 1/30 & 72 & 8.6(10.0) & 9.6(11.3) & 0.0(0.1) & $-$ & $-$ & $-$\\
& & & 102 & 9.8(11.3) & 10.9(12.8) & 0.1(0.1) & $-$ & $-$ & $-$\\ 
& & & 162 & 10.6(12.2) & 11.9(13.9) & 0.1(0.1) & $-$ & $-$ & $-$\\
SUBARU& $I$ & 1/30 & 70 & 4.7(5.4) & 5.5(6.4) & $-$  & $-$ & $-$ & $-$\\
& & & 90 & 5.8(6.5) & 6.9(7.8) &  0.1(0.1) &  $-$ & $-$ & $-$\\
& & & 120 & 6.1(6.9) & 7.3(8.4) &  0.1(0.1) &  $-$ & $-$ & $-$\\
& $R$ & 1/30 & 70 & 3.5(4.0) & 4.1(4.7) & $-$ & $-$ & $-$ & $-$\\
& & & 90 & 4.1(4.6) & 4.8(5.4) & $-$  & $-$ & $-$ & $-$\\
& & & 120 & 4.0(4.5) & 4.7(5.4) & $-$ & $-$ & $-$ & $-$\\
VISTAvis & $I$ & 1/30 & 60 & 34.5(39.9) & 39.2(46.3) & 0.3(0.3) & $-$ & $-$ & $-$\\ 
& & & 90 & 39.8(45.9) & 45.6(53.7) & 0.3(0.4) & $-$ & $-$ & $-$\\
& & & 150 & 43.2(50.0) & 49.7(58.7) & 0.4(0.5) & $-$ & $-$& $-$\\
& $R$ & 1/30 & 60 & 22.8(26.3) & 25.4(29.7) & 0.1(0.1) & $-$ & $-$ & $-$\\ 
& & & 90 & 25.3(29.0) & 28.4(33.1) & 0.1(0.2) & $-$ & $-$ & $-$\\ 
& & & 150 & 27.1(31.0) & 30.5(35.5)  & 0.2(0.2) & $-$ & $-$ & $-$\\
VISTA-IR & $J$ & 1/30 & 30 & 11.5(13.7) & 12.7(15.5) & 0.1(0.1) & $-$ & $-$ & $-$\\
& $H$  & 1/30 & 30 & 8.5(10.6) & 9.1(11.7) & 0.1(0.1) & $-$ & $-$ & $-$\\
& $K$  & 1/30 & 30 & 4.6(6.2) & 4.7(6.5) & $-$ & $-$ & $-$ & $-$\\

CFHT& $i'$ & 1/60 & 162 & 21.7(24.5) & 42.0(48.9) & 1.3(1.6) & $-$ & $-$ & $-$\\
SUBARU& $I$ & 1/60 & 120 & 7.5(8.3) & 15.2(17.2) & 0.6(0.7) &  $-$ & $-$ & $-$\\
VISTAvis& $I$ & 1/60 & 150 & 55.1(62.2) & 107.4(124.5) & 3.4(4.2) &  $-$ &  $-$ & $-$\\
VISTA--IR& $J$ & 1/60 & 30 & 15.2(17.6) & 28.3(33.9) & 0.7(0.9) & $-$ & $-$ & $-$\\
& $H$ & 1/60 & 30 & 11.9(14.4) & 20.9(26.3) & 0.5(0.7) & $-$ & $-$ & $-$\\
& $K$ & 1/60 & 30 & 7.0(9.1) & 11.3(15.2) & 0.3(0.4) & $-$ & $-$  & $-$\\

CFHT& $i'$ & 1/120 & 162 & 25.0(27.7) & 56.4(64.3) & 5.4(6.6) & 0.1(0.1) & 0.0(0.1) & $-$ \\
SUBARU& $I$ & 1/120 & 120 & 8.4(9.1) & 19.8(22.0) & 2.5(2.9) & $-$ & $-$ & $-$\\
VISTAvis & $I$ & 1/120 & 150 & 63.4(69.9) & 143.7(163.3)& 13.9(17.0)& 0.2(0.2) & 0.1(0.2) & $-$\\
VISTA--IR& $J$ & 1/120 & 30 & 17.9(20.0) & 39.1(45.6) & 3.0(3.5) & $-$ & 0.1(0.2) & $-$\\
& $H$ & 1/120 & 30 & 14.7(17.1) & 30.1(36.7) & 2.2(2.7) & $-$ & 0.1(0.1) & $-$\\
& $K$ & 1/120 & 30 & 9.3(11.7) & 17.2(22.6) & 1.3(1.7) & $-$ & $-$ & $-$\\

CFHT& $i'$ & 1/180 & 162 & 26.3(28.8) & 61.0(69.0) & 7.8(9.5) & 0.1(0.1) & 0.0(0.1)& $-$\\
SUBARU& $I$ & 1/180 & 120 & 8.8(9.4) & 21.2(23.3) & 3.6(4.2) & 0.1(0.1) & 0.0(0.1) & $-$ &\\
VISTAvis & $I$ & 1/180 & 150 & 66.6(72.6) & 155.3(175.0) & 20.1(24.5) & 0.3(0.4) & 0.1(0.3) & 0.0(0.1)\\
VISTA--IR& $J$ & 1/180 & 30 & 18.9(20.9) & 42.6(49.1) & 4.3(5.1) & 0.1(0.1) & 0.2(0.3) & $-$\\
& $H$ & 1/180 & 30 & 15.8(18.1) & 33.3(40.2) & 3.1(3.8) & $-$ & 0.1(0.2) & $-$\\
& $K$ & 1/180 & 30 & 10.4(12.7) & 19.6(25.4) & 1.9(2.4) & $-$ & 0.0(0.1) & $-$\\

CFHT& $i'$ & 2/120 & 162 & 28.7(30.8) & 70.2(77.8) & 16.7(20.2)&
0.4(0.5)& 0.1(0.2)& $-$ \\
SUBARU& $I$ & 2/120 & 120 & 9.4(9.9) & 23.7(25.7) & 7.4(8.6) & 0.2(0.3) & 0.1(0.1) & $-$ \\
VISTA--Vis & $I$ & 2/120 & 150 & 72.4(77.4) & 178.1(196.5) & 42.9(51.6) & 1.1(1.3) & 0.3(0.6) & 0.1(0.1)\\
VISTA--IR & $J$ & 2/120 & 30 & 20.9(22.5) & 50.1(56.1) & 9.4(11.0) & 0.2(0.2) & 0.3(0.6) & 0.0(0.1)\\
& $H$ & 2/120 & 30 & 18.0(19.9) & 40.9(47.6) & 7.1(8.5) & 0.1(0.1) & 0.2(0.4) & $-$\\
& $K$ & 2/120 & 30 & 12.6(15.0) & 25.7(32.3) & 4.3(5.4) & 0.1(0.1) & 0.1(0.1)  & $-$  \\

CFHT& $i'$ & 3/120 & 162 & 30.5(32.1) & 77.0(83.8) & 24.8(29.5) & 1.1(1.3)& 0.2(0.4) & 0.1(0.1)\\
SUBARU & $I$ & 3/120 & 120 & 9.8(10.2) & 25.5(27.2) & 10.6(12.2) &
0.5(0.6) & 0.1(0.3) & 0.0(0.1)\\
VISTAvis & $I$ & 3/120 & 150 & 76.7(80.7) & 194.7(211.1) & 63.6(75.2) & 2.8(3.3) & 0.5(1.2) & 0.1(0.3)\\
VISTA--IR & $J$ & 3/120 & 30 & 22.2(23.5) & 55.5(60.8) & 14.0(16.1) & 0.5(0.5) & 0.6(1.0) & 0.1(0.2)\\
& $H$ & 3/120 & 30 & 19.6(21.2) & 46.7(52.9) & 10.7(12.7) & 0.3(0.3) & 0.4(0.7) & 0.1(0.1)\\
& $K$ & 3/120 & 30 & 14.6(16.7) & 31.2(38.0)& 6.6(8.3)& 0.2(0.3) & 0.1(0.2) & $-$ \\

CFHT & $i'$ & 4/120 & 162 & 31.5(32.9) & 81.1(87.1) & 30.7(36.0) & 1.8(2.1) & 0.3(0.6) & 0.1(0.1)\\
SUBARU & $I$ & 4/120 & 120 & 10.0(10.3) & 26.6(28.0) & 12.9(14.6)  & 0.9(1.0) & 0.2(0.4) &  0.1(0.1)\\
VISTAvis & $I$ & 4/120 & 150 & 79.1(82.4) & 204.8(219.3) & 78.5(91.7) & 4.6(5.5) & 0.8(1.8) & 0.2(0.4)\\
VISTA--IR& $J$ & 4/120 & 30 & 23.0(24.0) & 58.8(63.4) & 17.2(19.7) & 0.8(0.9) & 0.8(1.5) & 0.2(0.2)\\
& $H$ & 4/120 & 30 & 20.6(21.9) & 50.5(56.1) & 13.4(15.7) & 0.5(0.6) & 0.6(1.0) & 0.1(0.2) \\
& $K$ & 4/120 & 30 & 15.8(17.8) & 35.1(41.8) & 8.4(10.4) & 0.4(0.4) & 0.2(0.3) & 0.0(0.1) \\
\hline
\end{tabular}
\label{tab:ret}
\end{table*}

\begin{figure}[t!]
\centering                     
\includegraphics[width=9.1cm]{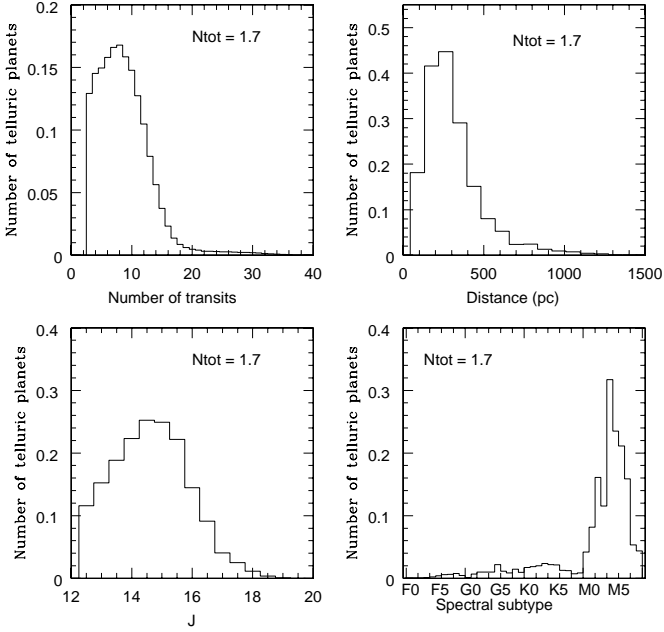}
\caption{Distribution of predicted  telluric  planet discoveries 
  ($\beta = 7$) for VISTA--IR  and the 4-years  $J$--band survey, as a
  function of   the number of  observed  transits (\emph{upper left}),
  distance (\emph{upper  right}),  magnitude (\emph{bottom left})  and
   of the host star spectral subtype (\emph{bottom right}).}
\label{fig:tvi}
\end{figure}

\begin{figure}[h!]
\centering                     
\includegraphics[width=9.1cm]{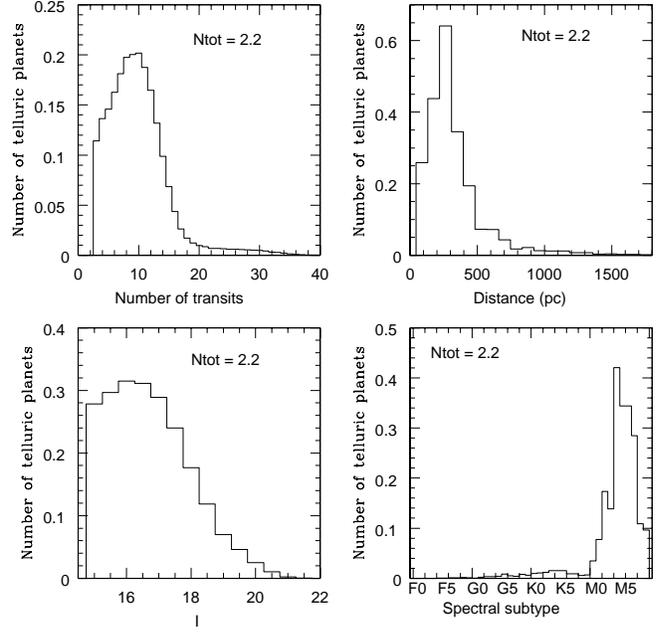}
\caption{Same as Fig.~\ref{fig:tvi} but for VISTA--Vis and 4 years of 
  observation through the $I$ filter.}
\label{fig:tvv}
\end{figure}

\subsubsection{Results}

In the following, we comment on the results presented in Table~\ref{tab:ret}.

\begin{itemize}  
  
\item   {\bf Optical  filters:}   optical surveys are   about twice as
  efficient in the $I$--band as in the  $R$--band.  This is mainly due
  to  the increased brightness of most $LMSS$  in   the   $I$  filter,  
  which  more   than compensates    for the lower  throughput  of  the  
  instruments in the $I$-band.   For this reason,  we   consider  only 
  $I$-band searches for surveys longer than 1 month. 
  
\item  {\bf Near-IR filters:} infrared   surveys are more efficient in
  the $J$--  and $H$--bands than in the  $K$--band.  In  the infrared,
  the gain in the number of $LMSS$ observed in $K$ does not compensate
  for the lower sensitivity of the detectors  at this wavelength.  The
  $J$ filter appears to be the best choice.
  
\item {\bf Exposure times:} in  the  optical,   the   longer  exposure 
  times do  not  always yield   to  larger  harvests.  Although   this 
  increases the $SNR$ across the duration of  the  transit,  it   also 
  leads  to  more  saturated stars,  so  that  the improvement in  the 
  harvest is negligible.
  
\item  {\bf (V)HJ vs.   IZ   planets:} even  for  the  shortest survey
  duration (30  days) and for the  least favorable detection criterion
  ($\beta=9$), the harvests in $(V)HJ$  are always large, with tens of
  discoveries. Increasing the duration of the survey does not increase
  very much the number   of $VHJ$ detected, but  increases drastically
  the number of giant planets found  in the $HZ$ and  in the $IZ$, the
  largest gain being for surveys longer than one year.  This is simply
  due to the  better  matching of the time  base-line  with the longer
  revolution periods of planets in the $HZ$ and the $IZ$.
  
\item {\bf Field vs. depth in the optical:} among the existing optical
  instruments,  the  harvests  obtained with  SUBARU  are smaller than
  those  obtained at the CFHT,  despite the larger telescope size: the
  improved depth  of SUBARU does not  compensate for the smaller field
  of view.
  
\item  {\bf Telluric  planets:} Under  the assumptions  used here, our
  fictitious   ground-based surveys seem  to have  a low potential for
  discovering  $TIZ$. VISTA--Vis has the   highest potential for  this
  purpose,  followed by VISTA--IR. Habitable  planets  seem out of the
  reach  of ground-based surveys. The  telluric planets which could be
  discovered  by VISTA--Vis or VISTA--IR   would probably orbit around
  M-dwarfs  relatively  close to the   Sun (see Figs.~\ref{fig:tvi} and
  \ref{fig:tvv}).

\end{itemize}

\begin{figure}
\centering                     
\includegraphics[width=9.1cm]{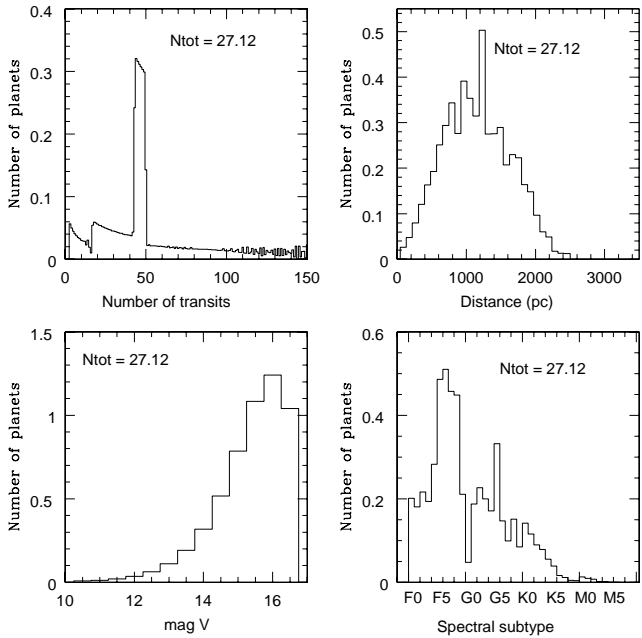}
\caption{Distribution of predicted planets (all types included)  
  discoveries using $\beta =  7$ for the  COROT mission, as a function
  of the number  of observed  transits (\emph{upper  left}),  distance
  (\emph{upper right}), magnitude (\emph{bottom left}) and of the host
  spectral subtype (\emph{bottom right}).}
\label{fig:coa}
\end{figure}

\subsection{Expected  harvest of the COROT mission}

COROT  (COnvection, ROtation and  planetary Transit) will be the first
satellite  launched with the   aim of detecting  exoplanets using  the
transit method (\cite{Rouan01}).  Another  goal of this mission is  to
carry out several   projects in asteroseismology  (\cite{Baglin}).  It
uses a  27 cm telescope  in combination with  two 2048$\times$2048 CCD
cameras, and a final field  of view of  3.5 deg$^{2}$.  During the 2.5
years of the mission, 5 fields in the  galactic plane will be observed
continuously, each   one  during 150   days.   The exoplanet detection
capability of this mission  has  already been analysed  (\cite{Borde})
but, since our  goal is a comparison  between surveys, we analyse  the
planet harvest  expected from COROT  using the same assumptions as for
the other surveys considered.

COROT   will  observe  fields  at  low  galactic   latitudes, i.e., in
directions with  a high density  of $LMSS$.   Indeed, each  field will
contain up to 12,000 dwarf stars with visual magnitudes between $V=11$
and $V=16.5$.   The exposure  time will be  31.7 seconds,  without any
filter.

\begin{figure}
\centering                     
\includegraphics[width=8.0cm]{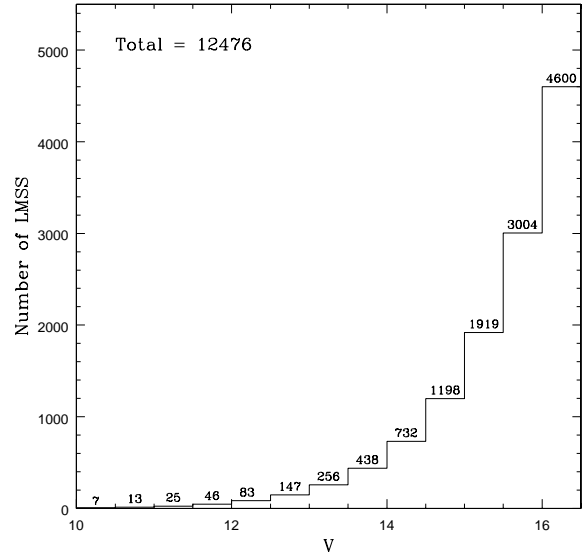}
\caption{Magnitude distribution of the $V \le 16.5$ $LMSS$ 
  in a typical COROT field.}
\label{fig:cos}
\end{figure}

We use the  expected number of photoelectrons  per  exposure for every
magnitude bin between $V=11$ and   $V=16.5$ as well as the   different
noise     contributions       given      on      the    COROT      web
site\footnote{http://www.astrsp-mrs.fr/projets/exoplan/corot/}  and in
Bord\'e (2003) to compute the theoretical $SNR$ per exposure. Using in
addition   the  duration of    one   field   observation  (150    days
continuously), and multiplying the obtained harvest by  5 to take into
account the 5 fields  that will be  monitored, we compute the harvests
presented   in Table~\ref{tab:cor}.  The  distributions  of  the total
number of  planets detected  for $\beta =   7$ as  a function  of  the
characteristics  of  the host stars  and  of   the number of  transits
observed  are shown    in Fig.~\ref{fig:coa}.  Fig.~\ref{fig:cos} shows 
the distribution of the $LMSS$ as afunction  of their magnitude in the  
$V$-band, for  a typical field observed by COROT.

\begin{table}
\centering
\caption{Results obtained for COROT, integrated 
over the duration of the whole mission.}
\begin{tabular}{ccccc}
\hline\hline $\beta$&$VHJ$&$HJ$&$GIZ$&$TIZ$\\
\hline
9 & 6.46 & 13.94 & 1.78 & 0.01\\ 
7 & 7.59 & 17.24 & 2.27 & 0.02\\
\hline
\end{tabular}
\label{tab:cor}
\end{table}

%%%%%%%%%%%%%%%%%%%%%%%%

\begin{figure}[t!]
\centering                     
\includegraphics[width=9.1cm]{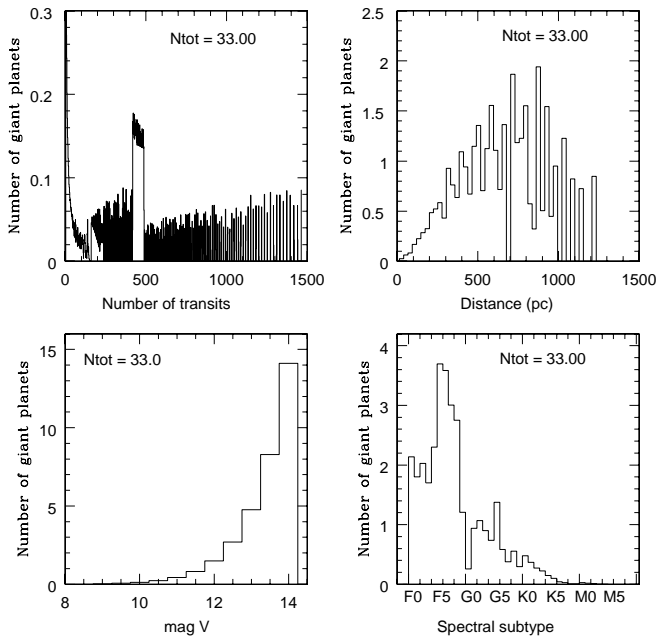}
\caption{Distribution  of  
  predicted giant planet discoveries using $\beta  = 7$ for KEPLER, as
  a  function of the number of  observed transits (\emph{upper left}),
  distance  (\emph{upper  right}), magnitude  (\emph{bottom left}) and
  host star spectral subtype (\emph{bottom right}) .}
\label{fig:kea}
\end{figure}

\begin{figure}[h!]
\centering                     
\includegraphics[width=9.1cm]{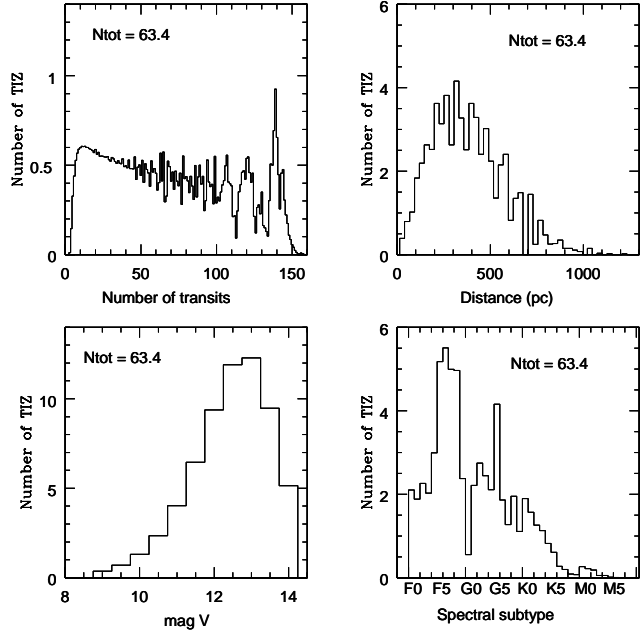}
\caption{Same as Fig.~\ref{fig:kea} but for 
  $TIZ$ ($\beta  = 7$).}
\label{fig:keb}
\end{figure}

Bord\'e et al. (2003)  estimated  that COROT  will  be able to  detect
numerous giant planets, a prediction that is confirmed by our results.
 It is nonetheless immediately   apparent that the expected   number of
giant planets  discoveries ($\sim$ 25) after  2.5 years is well within
the  reach  of   30 days ground-based  surveys. Only SUBARU would need 
a  survey of several months to be able to compete, due to its  smaller 
field of view.

\begin{figure}
\centering                     
\includegraphics[width=9.1cm]{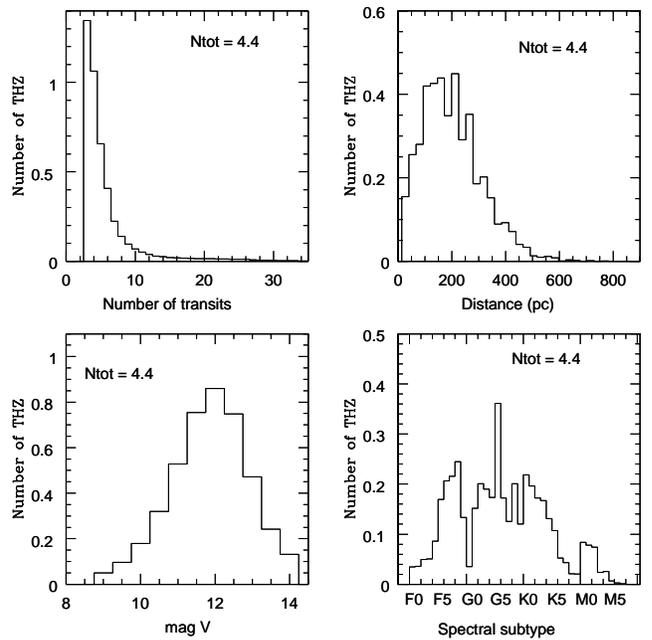}
\caption{Same as Fig.~\ref{fig:kea} but for $THZ$ ($\beta = 7$).}
\label{fig:kec}
\end{figure}

\begin{table}
\centering
\caption{Results obtained for the KEPLER mission.}
\begin{tabular}{ccccccc}
\hline\hline $\beta$&$VHJ$&$HJ$&$GIZ$&$GHZ$&$TIZ$&$THZ$\\
\hline
9 & 6.4 & 17.9 & 7.7 & 1.0 & 43.9 & 2.8\\
7 & 6.4 & 17.9 & 7.7 & 1.0 & 63.4 & 4.4\\
\hline
\end{tabular}
\label{tab:kep}
\end{table}

Bord\'e et  al. (2003) argue that COROT  will be able to  detect large
terrestrial planets   ($R \ge R_{\oplus}$,  with $R_{\oplus}$  = earth
radius)   in  close  orbits, provided  that   this  kind of  object is
frequent.  Under our assumptions, however,  the probability that COROT
finds a telluric planet is negligible,    even though we only consider
telluric planets similar  to the ones in our  Solar  System. 

\subsection{Expected harvest of the KEPLER mission}

The KEPLER mission is dedicated to the detection of earth-like planets
orbiting sun-like stars in the $HZ$  (\cite{Koch1}). This satellite is
scheduled for launch in 2007, and will observe continuously one single
105  deg$^{2}$ field during 4  years.  It  will monitor $\sim 100,000$
main sequence A-K stars with a mean  magnitude of $V=14$ using a 0.95 m
telescope and 3 seconds exposures.  Using  the technical data available
on  the   KEPLER web site\footnote{http://www.kepler.arc.nasa.gov}, we
compute the   $SNR$ for an individual exposure   time, and use   it to
estimate the planet harvest of KEPLER under the same assumptions as in
the  previous sections.    KEPLER will observe   a field  with a  mean
galactic latitude   of $13.3^{\circ}$, leading  to  a maximal distance
$D_L\sim$  1300 pc  according  to  Eq.~\ref{eq:a9}.   The  results  are
presented  in   Table~\ref{tab:kep}.            Figures~\ref{fig:kea},
\ref{fig:keb} and \ref{fig:kec}   show  the distribution of   expected
planet discoveries ($\beta=7$) as a function of the number of observed
transits and of the characteristics of their host star.

\begin{figure}
\centering                     
\includegraphics[width=8.0cm]{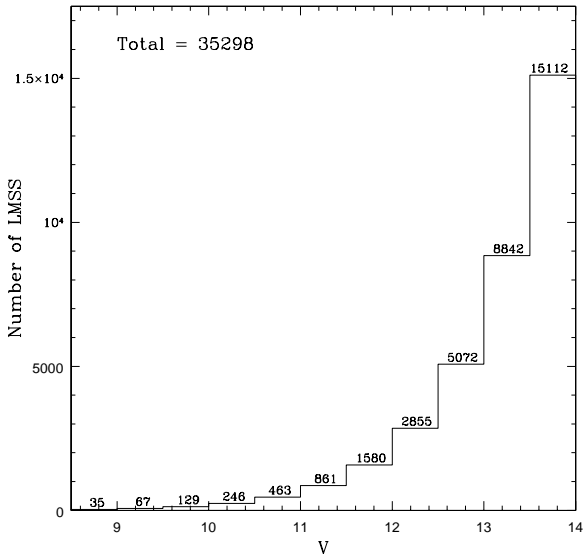}
\caption{Magnitude distribution of the $V \le 14$ $LMMS$ in the 
KEPLER field of view.}
\label{fig:kes}
\end{figure}

KEPLER  should be able to  detect  many telluric planets,  even in the
$HZ$. In contrast, the expected  harvest in giant planets is smaller
than the  one expected  for 30-days surveys carried out with VISTA--Vis. 
This is due  to the larger  field of view of KEPLER and to its shorter 
exposure times.  This strategy is certainly the   best one for hunting 
telluric  planets, but does not allow to observe enough $LMSS$ to detect  
a large number of giant planets in close orbits.

According to the KEPLER web site, the expected results of the mission
are the following:
\begin{itemize}
\item assuming that most telluric planets have  $R \sim 1 R_{\oplus}$,
  50 discoveries should be made.   This number increases  to 185 if most
  telluric planets have $R \sim 1.3 R_{\oplus}$ and  to 640 if most of
  them have $R \sim 2.2 R_{\oplus}$.
\item 135 inner-orbit giant planets and  30 outer-orbits giant planets
  should be discovered by KEPLER, plus  another 870 $(V)HJ$ discovered
  through the modulation of their reflected light.
\end{itemize}

These expectations are based on the following assumptions :
\begin{itemize}
\item The  number of main--sequence stars  monitored is $\sim 100,000$,
after the exclusion of the most active dwarf stars.
\item The variability of these $\sim  100,000$ stars on the time--scale
  of a transit is solar or close to solar.
\item All stars  host in average two  earth--size planets in the
  region between 0.5 and 1.5 AU.
\item Every star   host one single  giant planet in  a  jovian--like
  orbit.
\item On average, 1\% of the target stars  have a giant planet with an
  orbital period smaller than 1 week, 1\% with  periods between 1 week
  and 1 month and 1\% with a period between 1 month and 1 year.
\end{itemize} 

Detections  around  binaries  are   taken    into account  in    these
expectations.  The   detection criterion used   by  the KEPLER team is
$N_{min}=3$, $\beta=8$.

Our estimate of  the number of $LMSS$ in  the KEPLER field leads  to a
value of $\sim 35,000$.   Fig.~\ref{fig:kes} shows the distribution of
these   $LMSS$ as  a   function of   their  apparent   magnitude.  The
discrepancy between our  number of target stars  and the one from  the
KEPLER team may  have two origins: (1) we  do not take into account the
spectral types O, B  and A, (2) we assume  a constant density of $LMSS$
in a given  volume,  and  nothing outside,  (3) the  actual  extinction
coefficient $A_V$ of  the KEPLER field should  be  somewhat lower than
0.7 mag/kpc.

For the giant planets,  scaling our  result (the  sum  of the first  4
columns in Table~\ref{tab:kep},   $\sim 33$ planets)  to the   number of
targets stars expected by KEPLER  and taking into account the possible
detection around binaries would lead to a value in good agreement with
the  one  of  the  KEPLER team.  

Our assumptions about  telluric planets are  quite different from  the
ones used by the KEPLER team,  but our results  ($\sim$ $50 - 80$) are
in good  agreement, provided that  the values presented on  the KEPLER
web site include all telluric planets, and not only habitable ones.

KEPLER is a key project in the search  for  life  outside  our  Solar 
System. We can remark from Fig.~\ref{fig:kec}  that habitable planets 
would be detected mainly around G and K dwarfs  located about  200 pc   
from  the Sun, thus  around  nearby   solar--type stars. The predicted 
number of $THZ$ ($2 - 5$) is  small.  It could become  much higher if  
our radii distribution proved to be too pessimistic. 

\section{Discussion - Conclusions}

The main purpose of our simulations was to  weight the advantages and 
disadvantages  of  various   ground--  and  space--based  searches  for 
exoplanets   using  the transit method. Tables 2 to 6  summarize  our 
results and provide the expected harvest in exoplanets  for a   broad  
variety  of telescope/instrument combinations.  Our  main conclusions 
are the following:

\begin{enumerate}
  
\item As far as telluric planets in the Habitable Zone are concerned,
  space--based surveys are the   only   viable  option. Such  searches 
  remain   extremely   difficult.  They  not  only  require  a  space 
  instrument, but  also a very wide field of view.  From space,  only 
  the KEPLER mission  should be able to find  telluric planets in the
  Habitable Zone.
  
\item Telluric  planets in the   Intermediate Zone are much easier to
  discover. KEPLER could detect more than  40 of them during its four
  years of observations. Only a few   (1--2) $TIZ$ might be discovered
  from ground--based surveys  of the same duration,  using  VISTA--Vis.
  
\item Ground--based searches are better than  space searches at finding
  giant planets. While KEPLER is about as efficient as CFHT at finding
  giant planets  in  the  Habitable  Zone (with 1   expected discovery
  \emph{vs} 2), a CFHT search is 4 times better than KEPLER at finding
  the same  planets in the  Intermediate Zone, and  5 times better for
  $(V)HJ$. This is due to the much deeper exposures.

\end{enumerate}

Ground--based  and  space--based transit  searches   are complementary.  
Because   they  go deeper, ground--based    searches  easily find large
planets with a short period,  such as $(V)HJ$.  Space searches  remain
mandatory for telluric   planets. COROT  will be a pionner mission in
this field, but we   shall  have to   wait for  KEPLER to  obtain  a
significant harvest of such objects,
provided that they are not significantly less common than expected.\\

The above results give orders of  magnitude estimates for the expected
harvests and allow  to emphasize the relative  merits and drawbacks of
the different  searches.  However, a  word of caution should be given,
to  avoid over--interpretation  of the  results.  One need to  be aware
that:

\begin{itemize}
  
\item Space  missions  like COROT or KEPLER   will defocus the images,
  increasing drastically the size of the  Point Spread Function (PSF). 
  While this will minimize  saturation of bright objects and  increase
  the $SNR$ per image, it  might result in  severe image blending.  The
  corresponding loss of efficiency in transit detection will depend on
  the  method used  to post--process  the data,   as do the  effects of
  blends.   This  is why we have   deliberately chosen not to  take PSF
  convolution into consideration and  to leave it for  a future  work. 
  All our estimates  are therefore {\it  upper limits on the  expected
    harvests.}
  
\item The weather simulations used are very simple, and    could  be 
  somewhat optimistic. Real weather conditions could   lead to lower
  harvests for ground---based surveys.
  
\item The   photometric  techniques  required  for  the  analysis of 
  ground--based surveys are very efficient in the  optical (e.g., PSF 
  fitting,  image subtraction, image deconvolution), but may be less
  efficient  for near--IR  data, where  the sky subtraction   is more 
  critical. We  therefore expect our near--IR harvest estimates to be 
  slightly more optimistic  than   the  optical  estimates. 

\item In the near--IR, second--order extinction effects can have a 
  large impact on the photometric accuracy (\cite{Bailer-Jones}). Time 
  dependent atmospheric extinction depends on the spectral energy
  distribution of the target. This may imply the need to use reference 
  objects of the same spectral type as the target stars and complicate 
  further the analysis of near--IR data.

\end{itemize}

\begin{acknowledgements} 
  The  authors would like  to   thank G.  Mall\'en--Ornelas and T. Guillot
  for useful discussions and suggestions,  and A.  Udalski for  providing 
  informations about OGLE-III. We acknowledge financial support  from  
  the Prodex-ESA Contract 15448/01/NL/Sfe(IC).
\end{acknowledgements}

\end{document}